\documentclass[aps,prb,preprint,groupedaddress]{revtex4}

\usepackage{graphicx}
\usepackage{latexsym}
\usepackage{amsmath,amssymb}

\begin{document}

\title{
Deconfinement criticality 
for the spatially anisotropic
triangular antiferromagnet with the ring exchange
}

\author{Yoshihiro Nishiyama}
\affiliation{Department of Physics, Faculty of Science,
Okayama University, Okayama 700-8530, Japan}

\date{\today}

\begin{abstract}
The spatially anisotropic triangular antiferromagnet
is investigated with the numerical diagonalization method.
As the anisotropy varies,
the model changes into a variety of systems such as
the one-dimensional, triangular,
and square-lattice antiferromagnets.
Taking into account such a geometrical character,
we impose the screw-boundary condition, which interpolates
smoothly the one- and two-dimensional lattice structures.
Diagonalizing the finite clusters with $N=16,20,\dots,32$ spins,
we observe
an intermediate phase between
the VBS and N\'eel phases.
Suppressing the intermediate phase by applying the ring exchange,
we realize a direct VBS-N\'eel transition.
The simulation data indicate that the transition is a continuous one
with the correlation-length critical exponent $\nu=0.80(15)$.
These features are in agreement with 
the deconfinement-criticality scenario advocated by
Senthil and coworkers in the context of the high-temperature superconductivity.
\end{abstract}


\maketitle


\section{\label{section1} Introduction}

According to the deconfinement-criticality scenario,\cite{Senthil04a,Senthil04b,Senthil05,Alicea05}
in $(2+1)$ dimensions,
the phase transition separating
the VBS and N\'eel phases is
continuous, accompanied with unconventional critical indices.
Naively, such a transition should be discontinuous,\cite{Senthil04b}
because the adjacent phases 
possess distinctive order parameters
such as the VBS-coverage pattern, and the sublattice magnetization, respectively.
A good deal of field-theoretical investigations
have been made
to clarify this issue.\cite{Senthil06,Tanaka05}
For instance, as a low-energy effective description,
the QED${}_3$ theory has been investigated;\cite{Liu05,Dellenschneider06,Ghaemi06}
it would be intriguing that 
the theory exhibits a deconfinement transition.\cite{Thomas07}
On the one hand, because of the magnetic frustration,
the Monte Carlo simulation suffers from the
negative-sign problem to realize the VBS phase.
However,
in recent Monte Carlo simulations,\cite{Sandvik07,Melko08}
the biquadratic interaction, rather than the magnetic frustration,
has been utilized.
Thereby, it turned out 
that the biquadratic-interaction-driven
transition is a continuous one with unconventional
critical indices.
(On the contrary, in Refs. \onlinecite{Jian08,Kruger06,Kuklov08},
it was claimed that the transition would be 
a weak first-order one.)

In this paper, 
we investigate
the spatially anisotropic triangular antiferromagnet\cite{Coldea01,Yunoki04}
with the ring exchange
by means of the numerical diagonalization method.
As the anisotropy varies,
the model changes into the one- and two-dimensional systems,
and correspondingly, a variety of phases appear.
To cope with such a geometrical peculiarity,
we impose the screw-boundary condition,
which interpolates the one- and two-dimensional
lattice structures smoothly.

To be specific,
we present the Hamiltonian for
the spatially anisotropic triangular
antiferromagnet with the ring exchange;
\begin{equation}
\label{Hamiltonian}
{\cal H}=
J \sum_{\langle ij \rangle} {\bf S}_i \cdot {\bf S}_j
+J'\sum_{\langle \langle ij \rangle \rangle}
                 {\bf S}_i \cdot {\bf S}_j
+J_4 \sum_{[ijkl]}  (P_{ijkl} +P^{-1}_{ijkl})
           .
\end{equation}
The quantum spin-$1/2$ operators $\{ {\bf S}_i \}$ are
placed at each triangular-lattice point $i$.
The symbol $P_{ijkl}$ denotes a ring-exchange operator
with respect to a plaquette $[i,j,k,l]$ consisting of two adjacent triangles;
namely, as to a plaquette state 
$| {}^{S_1}_{S_3}\Box^{S_2}_{S_4} \rangle$,
the operation $P_{1234}$ translates it into
$| {}^{S_3}_{S_4}\Box^{S_1}_{S_2} \rangle$.
The summations
$\sum_{\langle ij \rangle}$,
$\sum_{\langle \langle ij \rangle \rangle}$, and
$\sum_{[ijkl]}$
run over all possible vertical nearest-neighbor pairs,
remaining nearest-neighbor pairs, and plaquette spins, respectively;
the triangular lattice is directed so that one of the triangular edges 
points upward.
The parameters 
$J$, $J'$, and $J_4$ are the corresponding coupling constants.
(In the next section, we present
an explicit expression for the Hamiltonian matrix,
referring to
the technical details of the screw-boundary condition.)
Hereafter, we consider $J$ as a unit of energy;
namely, we set $J=1$.

\begin{figure}
\includegraphics[width=100mm]{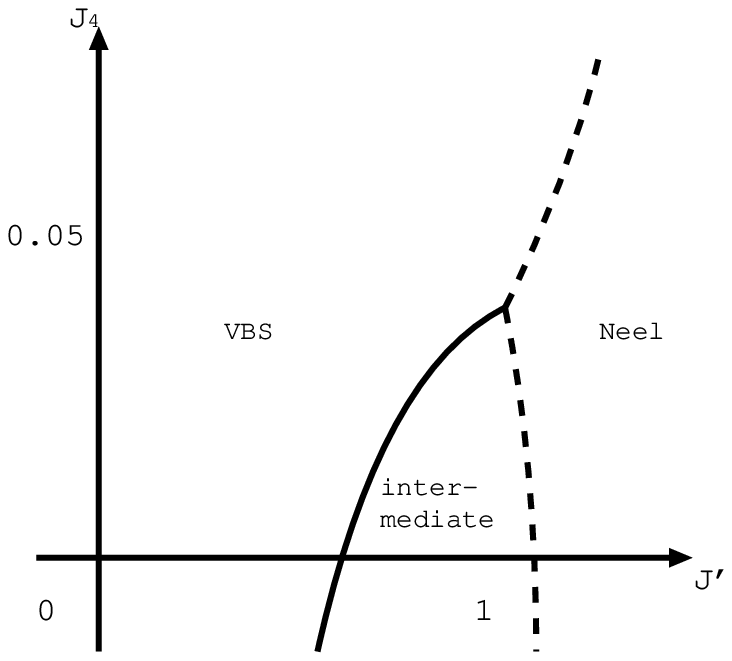}%
\caption{\label{figure1}
A schematic phase diagram for the spatially anisotropic 
triangular antiferromagnet (\ref{Hamiltonian}) is presented.
The solid (dashed) line stands for 
the first- (second-) order phase boundary.
Our concern is to realize a direct VBS-N\'eel transition
by applying the ring exchange $J_4$, and analyze the singularity
in the context of the deconfinement criticality.
Concerning the singularities of the phase boundaries surrounding the intermediate phase,
we follow the conclusion of Ref. \onlinecite{Weng06}.
}
\end{figure}

In Fig. \ref{figure1},
we present
a schematic phase diagram;
the details are explained in Sec. \ref{section3}.
As mentioned above,
the aim of this paper is to survey 
the direct VBS-N\'eel transition;
in this sense, the ring exchange $J_4$ is significant
to realize the VBS-N\'eel transition.
A number of limiting cases were studied in Refs. \onlinecite{Trumper99,Zheng99,Zheng06,Chung01,Weng06,Misguich99}:
First, the case $J_4=0$ was investigated
with the spin-wave,\cite{Trumper99} series-expansion,\cite{Zheng99,Zheng06}
large-$N$,\cite{Chung01} and numerical-diagonalization\cite{Weng06} methods.
The regime of the intermediate 
(triangular antiferromagnetic) phase\cite{Bernu94} was estimated
as 
$0.27 < J'<2$,
$0.25 < J'< 1.43$,
$0.13 < J'< 1.71$, and
$0.78(5) < J' < 1.15(10)$,
respectively.
(Some analyses predict two types of intermediate phases.
Such a detail is ignored for simplicity.)
These results appear to be unsettled.
It is a purpose of this paper to 
survey the intermediate phase.
Second, 
the spatially isotropic ($J'=J$) case
in the presence of
the ring exchange was investigated in Ref. \onlinecite{Misguich99};
here, 
the generic types of ring-exchange interactions were considered in the context of
the Helium adsorbate.
It was reported that the $J_4$-driven phase transition occurs 
in agreement with our observation.

As mentioned above, the model (\ref{Hamiltonian})
has a geometrical peculiarity.
That is,
as the spatial anisotropy $J'$ changes,
the model (\ref{Hamiltonian}) reduces to 
the one-dimensional ($J'=0$),
triangular ($J'=1$), and
square-lattice ($J'\to \infty$) antiferromagnets successively.
(Hence, for sufficiently large $J'$,
the conventional non-collinear N\'eel phase
appears.)
Notably enough,
the phase diagram, Fig. \ref{figure1},
reflects this geometrical character.
In order to take into account
this geometrical character,
we implemented
the screw-boundary condition,
which interpolates the one- and two-dimensional-lattice structures smoothly.

In fairness, it has to be mentioned that
the VBS-N\'eel transition was studied 
for 
the frustrated square-lattice antiferromagnet,
namely, the $J_1$-$J_2$ model.\cite{Oitmaa96,Sirker06,Poilblanc06}
According to the series-expansion method,\cite{Oitmaa96}
the N\'eel ($J_2/J_1 \lesssim0.4$),
VBS ($0.4 \lesssim J_2/J_1  \lesssim 0.6$), and collinear
($0.6 \lesssim J_2/J_1$) 
phases appear successively, as the magnetic frustration changes.
The VBS phase seems to be dominated by the presence of
the collinear phase.
(Note that for $J_2/J_1 \to \infty$, the system reduces to
two independent square-lattice antiferromagnets.
The collinear state
consists of two independent N\'eel orders.)
In this paper, we dwell on the triangular
antiferromagnet (\ref{Hamiltonian}),
which exhibits
an isolated VBS-N\'eel transition.

The rest of this paper is organized as follows.
In Sec. \ref{section2}, we explicate the simulation algorithm,
placing an emphasis on the screw-boundary condition.
In Sec. \ref{section3}, we show the finite-size-scaling analysis
of the simulation data.
In Sec. \ref{section4}, we present the summary and discussions.

\section{\label{section2}Simulation method: Screw-boundary condition}

In this section, we present an explicit expression for
the Hamiltonian, Eq. (\ref{Hamiltonian}),  under the screw-boundary condition.

To begin with, we present a schematic drawing of the finite-size cluster
in Fig. \ref{figure2}.
As shown in the figure,
the spins constitute a one-dimensional ($d=1$)
alignment $\{ {\bf S}_i \}$ ($i=1,2,\dots,N$).
The dimensionality is lifted to $d=2$ by the bridges over the $v$th-neighbor 
interactions.
As mentioned in the Introduction,
the spatially anisotropic triangular antiferromagnet
possesses a geometrical character such that
it
reduces to a one-dimensional antiferromagnet in the limit $J' \to 0$.
In this sense,
the geometrical peculiarity is seized by the screw-boundary condition.
Actually, for a rectangular cluster
with
the system size $6\times 6$, 
for instance,
the length of the independent chains in the limit $J' \to 0$
is merely $L=6$.
On the contrary, owing to the screw-boundary condition,
we attain treating
the chain length $L = 32$ along the $J$-bond direction.

To be specific,
we present an explicit expression for the Hamiltonian matrix.
We propose the following expression;
\begin{equation}
\label{actual_Hamiltonian}
{\cal H}=
J H(1)
+J'  ( H(v)+  H(v+1) )
+J_4 ( H_4 (1,v)+H_4(1,v+1)+H_4 (v+1,v) )    .
\end{equation}
Here, the $v$th-neighbor Heisenberg interaction $H(v)$ is given by
\begin{equation}
H(v)=\sum_{i=1}^{N} {\bf S}_i \cdot {\bf S}_{i+v} .
\end{equation}
(The periodic condition, namely, ${\bf S}_{N+i}={\bf S}_i$,
is imposed.)
Similarly, the ring exchange is introduced via
\begin{equation}
H_4(j,v) = \sum_{i=1}^{N} ( P_{i,i+j,i+v,i+j+v} +h.c.) .
\end{equation}
We set the screw pitch to
\begin{equation}
\label{screw_pitch}
v(N)=
\left\{
\begin{array}{ll}
n(\sqrt{N})+1 & for \  N \ge 24   \\
n(\sqrt{N})   &  otherwise
\end{array}
\right  .
\end{equation}
with the round-off function $n(x)=[x+0.5]$
and Gauss' symbol $[\dots ]$; {\it i.e.}, $n(2.4)=2$.
The screw pitch $v(N)$ converges to $v(N) /\sqrt{N} \to 1$ 
for large system sizes $N\to \infty$;
hence, the spins form a $\sqrt{N} \times \sqrt{N}$ network 
embedded on the torus.
The rule, Eq. (\ref{screw_pitch}), is intended to
suppress the finite-size errors;
actually, by Eq. (\ref{screw_pitch}),
we can set
the screw-pitch $v$ to an even number (for small $N$), which turns out to improve
the finite-size behavior even
for small system sizes.
More specifically,
the screw-boundary condition introduces a frustration particularly
for the N\'eel-type magnetism ($J'\to \infty$),
and the frustration effect is suppressed by the above rule, Eq. (\ref{screw_pitch}).



The above formulae complete the basis of our scheme.
As shown in Fig. \ref{figure2},
the embedding geometry under the screw-boundary
condition is essentially one-dimensional,
admitting us to calculate
the Hamiltonian-matrix elements
systematically
with Eq. (\ref{actual_Hamiltonian}).
In the next section, utilizing the Lanczos 
method, we diagonalize the Hamiltonian
matrix for the system sizes $N \le 32$.


\section{\label{section3}
Numerical results}

In this section, we present the numerical results.
We calculate
the excitation gap
\begin{equation}
\label{energy_gap}
\Delta E_i(k,S^z_{tot})=
    E_i(k,S^z_{tot})   - E_0(0,0^+)
                ,
\end{equation}
with
the $i$th low-lying energy 
 $E_i(k,S^z_{tot})$ 
($i=1,2,\dots$) within the sector
$(k,S^z_{tot})$.
Here,
the index $k$ 
denotes the wave number within
the Brillouin zone $-\pi \le k \le \pi$.
We impose the screw-boundary condition (Fig. \ref{figure2}),
and the Bloch wave $k$ extends along the spiral ($J$-bond) chain;
hence, the reciprocal space is one-dimensional. 
The quantum number $S^z_{tot}$ denotes an eigenvalue of the operator $\sum_{i=1}^N S^z_i$.
In the case of $S^z_{tot}=0$, additionally,
we introduce an index $\pm$,
which specifies the
inversion symmetry with respect to $S^z_i \to -S^z_i$.
The sector $(0,0^+)$ 
contains the ground state.
In this sector,
we shift the $i$ index so as to express the ground-state energy as $E_0(0,0^+)$
via $i \to i-1$.
(The ground-state energy is the starting point of all excitations,
and it is sensible to index the ground-state energy as $E_{i=0}$ rather than $i=1$.)
The linear dimension $L$ of the cluster is given 
by 
\begin{equation}
L  = \sqrt{N} ,
\end{equation}
because the $N$ spins constitute a two-dimensional network
as shown in Fig. \ref{figure2}.

\subsection{\label{section3_1}
Spatially anisotropic triangular antiferromagnet: $J_4 =0$}

In this section, we survey the regime without the
ring exchange $J_4=0$;
as mentioned in the Introduction,
this case has been studied
in Refs. \onlinecite{Trumper99,Zheng99,Zheng06,Chung01,Weng06},
and the details of the intermediate phase remain unclear.

In Fig. \ref{figure3},
we plot the excitation gap $\Delta E_1(\pi,0^+)$
for $J_4=0$, various $J'$ and $N=16,20,\dots,32$.
We notice that the level crossings take place at $J' \approx 0.65$ and $J'\approx 1.1$.
That is,
the softening instability, 
$\Delta E_1(\pi,0^+)  <0$, occurs in the intermediate regime.
We estimate the range of the intermediate phase as
\begin{equation}
\label{intermediate_phase}
 0.65(15) < J' < 1.1(1)
.
\end{equation}
Here, as an error indicator,
we utilize the data scatters
of the $J'$-intercept among
$N=20$, $24$, $28$, and $32$.
(Several related studies are overviewed afterward.)

Surveying various parameter ranges, we found that
the elementary-excitation gap opens at
either $k=0$ or $\pi$.
The softening of the branch $k=\pi$ suggests 
that the magnetic order along the $J$-bond direction
is unstable against a
staggered modulation.
Such a staggered modulation fits the boundary condition
(constraint)
such that 
the chain length $N$ is always set to an even number.
On the one hand, as shown in Fig. \ref{figure2},
the number of spiral turns, $N/v$, of the chain 
is a fractional number, and the
magnetism along the spiral direction may not fit the embedding geometry.
Hence, the staggered order along the chain direction
becomes even stabilized, resulting in the $k=\pi$
softening.
On the one hand,
in the
VBS phase $J'<0.65$, the energy gap $\Delta E_1(\pi,0^+)$
gets closed as the system size enlarges;
eventually, the ground state may be doubly degenerated
in the thermodynamic limit.
This double degeneracy suggests
that the $J$-bond chain is covered by the dimers.
(In this sense, the VBS picture of 
the present system is not so complicated,
as compared to that of the square lattice.\cite{Sindzingre04,Mambrini06})

On the contrary, in the N\'eel phase
$1.1 < J'$, a positive gap $\Delta E_1(\pi,0^+)>0$ starts to open.
In fact, in the limit $J' \to \infty$,
the model reduces
to the square-lattice antiferromagnet.
Hence,
the spins along the diagonal ($J$-bond) direction align ferromagnetically,
and the $k=\pi$ excitation exhibits a mass gap.

It is a good position to 
make an overview of the related studies.
According to 
the spin-wave,\cite{Trumper99} series-expansion,\cite{Zheng99} 
large-$N$,\cite{Chung01} and diagonalization studies,\cite{Weng06}
the range of the intermediate phase is estimated as
$0.27 < J'<2$,
$0.25 < J'< 1.43$,
$0.13 < J'< 1.71$, and
$0.78(5) < J' < 1.15(10)$,
respectively.
Our result, Eq. (\ref{intermediate_phase}),
indicates that the VBS phase persists up to a considerably large $J'$,
suggesting that 
the VBS phase is robust.
Similar conclusion was drawn from
the diagonalization study by Weng and coworkers.\cite{Weng06}
They diagonalized 
the
rectangular clusters
with the sizes $6\times4$, $8\times4$, and $6\times6$.
Such a rectangular geometry is
suitable for investigating the N\'eel-type magnetic structure.
On the contrary, the screw-boundary condition
meets the quasi-one-dimensional system (VBS phase).
The agreement between these approaches would be encouraging.

As a reference, in Fig. \ref{fig_new},
we present the ground-state energy
per unit cell, $E_0(0,0^+)/N$, with $N=32$ for the
same parameter range as that of Fig. \ref{figure3}.
In the small-$J'$ regime, the ground-state 
energy is close to the Bethe-ansatz solution, $E_0/N=-0.443\dots$,
for the one-dimensional Heisenberg antiferromagnet.
This fact suggests that the VBS phase is of one-dimensional
character.

Last, we mention a number of remarks concerning the phase diagram.
We made similar analyses for various values of $J_4 \sim 0$,
The result is summarized in Fig. \ref{figure1};
as suggested by Eq. (\ref{intermediate_phase}), the intermediate-phase boundaries
are not determined very precisely, and the boundaries in Fig. \ref{figure1} are 
only schematic.
(The critical branch separating the VBS and N\'eel phases
is considered in the next section.)
Concerning the singularities of the phase boundaries
surrounding the intermediate phase,
we followed the claim \cite{Weng06} that
the intermediate-VBS (-N\'eel) phase transition
is discontinuous (continuous).

\subsection{\label{section3_2}
Suppression of the intermediate phase by the ring exchange $J_4$}

In this section, we explore the regime
with the ring exchange $J_4 \ne 0$.

In Fig. \ref{figure4}, we plot the excitation gap 
$\Delta E_1  (\pi,0^+)$
for $J_4=0.07$, various $J'$ and $N=16,20,\dots,32$.
Comparing the result with that of Fig. \ref{figure3},
we notice that the ring exchange $J_4$ suppresses the
intermediate phase (softening instability).
As mentioned in the Introduction,
the suppression of the intermediate phase by $J_4$ was demonstrated
in Ref. \onlinecite{Misguich99}
at $J'=1$.
In the present study, we have yet another parameter $J'$,
and we are able to investigate the $J'$-driven
VBS-N\'eel transition.

In Fig. \ref{figure5},
we plot the scaled energy gap $L^z \Delta E_1 (0,1)$ 
for $J_4=0.07$, various $J'$, and $N=16,20,\dots,32$.
Note that the sector $S^z_{tot}=1$ corresponds to the triplet excitation
created preferentially on the $J$-bond chain.
The behavior of the triplet excitation contains information
on the phase transition from the VBS phase.
Here, we set the dynamical critical exponent
to $z=1$, following the conclusion of the Monte Carlo analyses\cite{Sandvik07,Melko08}
for the square-lattice antiferromagnet.
(Note that the energy gap $\Delta E_1$ is proportional to the reciprocal correlation length,
and the product $L \Delta E_1$ is a dimensionless (scale invariant) quantity.)
According to the scaling theory,
the curves of the scaled energy gap 
should intersect each other at the critical point.
In fact, we observe that a phase transition 
takes place at $J'=1.2$.
Surveying various parameter ranges, we arrive at the phase diagram,
as depicted in
Fig. \ref{figure1}.

A remark is in order.
As mentioned in the Introduction, naively,
the VBS-N\'eel transition should be discontinuous;\cite{Senthil04b}
actually,
the adjacent phases possess distinctive order parameters
such as the dimer-coverage pattern and the sublattice magnetization,
respectively.
However,
according to the deconfinement-criticality scenario,\cite{Senthil04a}
the spinon deconfining from a sea of singlet dimers 
changes the singularity
to a continuous one.
Our result supports this scenario.
In the next section, surveying a critical index, 
we investigate the criticality of the VBS-N\'eel transition more in detail.

\subsection{\label{sectioin3_3}Deconfinement criticality}

In this section,
we estimate the critical exponent $\nu$ for the VBS-N\'eel transition.

In Fig. \ref{figure6},
we present the finite-size-scaling plot, 
$(J'-J_c')L^{1/\nu}$-$L \Delta E_1(0,1)$,
for $J_4=0.07$, various $J'$, and $N=16,20,\dots,32$.
Here, we set the scaling parameters to $J_c'=1.2$ and $\nu=0.8$;
note that the former parameter $J_c'=1.2$ was determined in Fig. \ref{figure5}.
The data of Fig. \ref{figure6} collapse into a scaling curve, confirming that the
transition is indeed critical.
Moreover, the critical exponent 
acquires an enhancement, 
as compared to that of the $3d$ Heisenberg 
universality, $\nu=0.7112(5)$.\cite{Compostrini02}
(An overview of the related studies is addressed afterward.)
Similarly, in Fig. \ref{figure7},
we present the finite-size-scaling plot, 
$(J'-J_c')L^{1/\nu}$-$L \Delta E_1(0,1)$,
for $J_c'=1.4$, $\nu=0.8$, $J_4=0.1$, and $N=16,20,\dots,32$.
Again, the data collapse satisfactorily.
Surveying various parameter ranges,
we arrive at 
\begin{equation}
\label{exponent}
\nu =0.80(15)
                        .
\end{equation}

This is a good position to make an overview of the preceding 
Monte Carlo studies.
As for
the square-lattice antiferromagnet, 
the biquadratic-interaction-driven
VBS-N\'eel transition was investigated in
Refs. \onlinecite{Sandvik07} and \onlinecite{Melko08},
and the 
critical exponent was estimated as
$\nu=0.78(3)$ and $\nu=0.68(4)$, respectively.
Moreover, 
as for the quasi-one-dimensional
spin-1 antiferromagnet,
the index $\nu=1/2.9$ was reported;\cite{Harada07} 
see Ref. \onlinecite{Nogueira07} for a field-theoretical interpretation.
(Note that these models are free from the negative-sign problem,
and the quantum Monte Carlo method is applicable.)
We notice that the results are not quite settled. 
A notable point is that the 
exponent\cite{Sandvik07} $\nu=0.78(3)$ is significantly larger than that of the
$d=3$ Heisenberg
universality class $\nu=0.7112(5)$,\cite{Compostrini02} suggesting a peculiarity of the deconfinement criticality.
Our result, Eq. (\ref{exponent}), also suggests a tendency of an enhancement as to $\nu$.
Nevertheless, 
our simulation result provides an evidence that the VBS-N\'eel transition
is a critical one
in agreement with the deconfinement-criticality
scenario advocated by Senthil and coworkers.

\section{\label{section4}Summary and discussions}

\begin{figure}
\includegraphics[width=100mm]{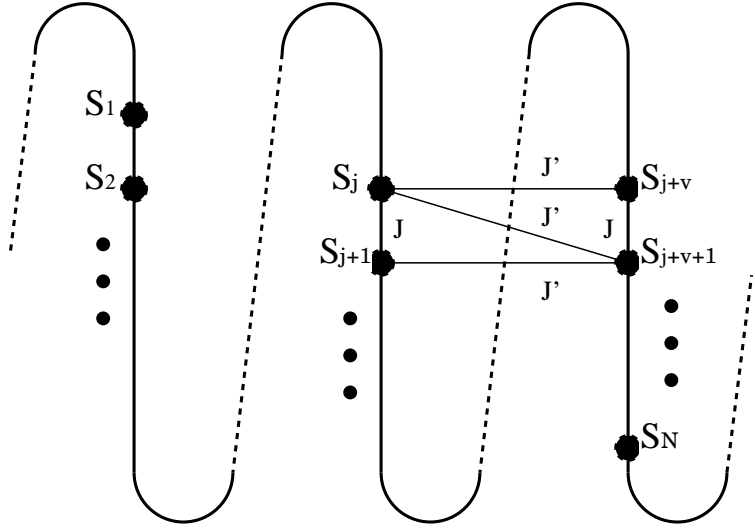}%
\caption{\label{figure2}
A schematic drawing of the spin cluster 
under the screw-boundary condition for the
spatially anisotropic triangular antiferromagnet 
(\ref{Hamiltonian}) is presented.
As indicated,
the spins constitute a one-dimensional ($d=1$) alignment
$\{ S_i\}$
($i=1,2,\dots,N$) via the longitudinal coupling $J$.
The dimensionality is lifted to $d=2$ by the bridges over the
$v$th-neighbor pairs through the transverse coupling $J'$.
The screw pitch $v$ is given by Eq. (\ref{screw_pitch}).
}
\end{figure}

\begin{figure}
\includegraphics[width=100mm]{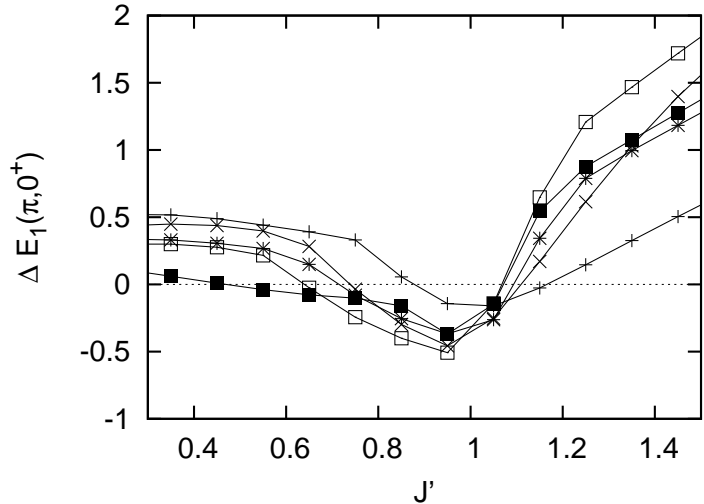}%
\caption{\label{figure3}
The energy gap $\Delta E_1(\pi,0^+)$,
Eq. (\ref{energy_gap}), is plotted for $J_4=0$, various $J'$, and 
$N=$
($+$) 16,
($\times$) 20,
($*$) 24,
($\Box$) 28, and
($\blacksquare$) 32.
A softening instability, $\Delta E_1<0$, takes place 
in the intermediate phase
$0.65(15)<J'<1.1(1)$.
}
\end{figure}

\begin{figure}
\includegraphics[width=100mm]{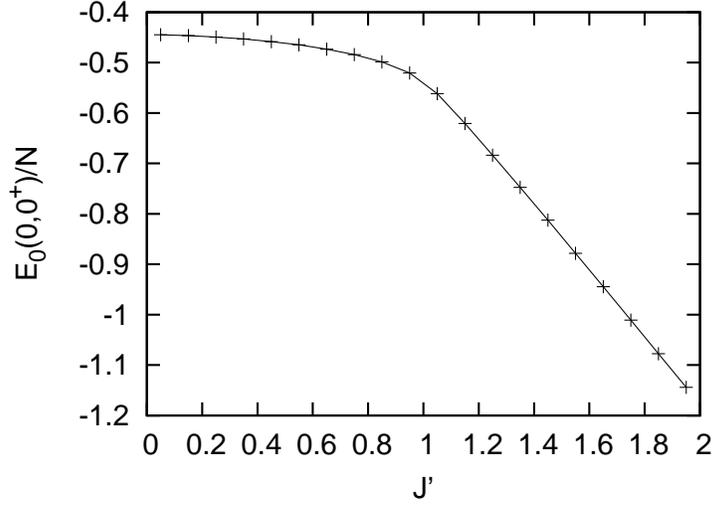}%
\caption{\label{fig_new}
The ground-state energy per unit cell, $E_0(0,0^+)/N$,
with $N=32$
is presented for
the same parameter range as that of Fig. \ref{figure3}.
In the small-$J'$ regime, the ground-state
energy is close to the Bethe-ansatz solution,
$E_0/N=-0.443\dots$, for the one-dimensional Heisenberg antiferromagnet.
}
\end{figure}

\begin{figure}
\includegraphics[width=100mm]{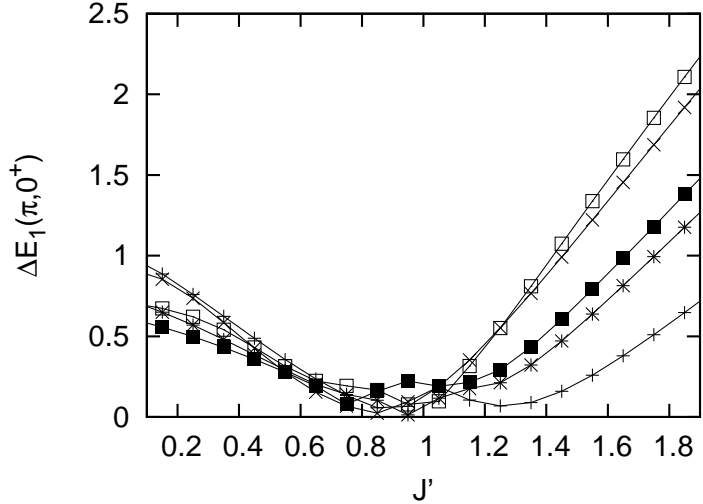}%
\caption{\label{figure4}
The energy gap $\Delta E_1(\pi,0^+)$,
Eq. (\ref{energy_gap}), is plotted for $J_4=0.07$, various $J'$, and 
$N=$
($+$) 16,
($\times$) 20,
($*$) 24,
($\Box$) 28, and
($\blacksquare$) 32.
Owing to the ring exchange,
the softening instability, namely, $\Delta E_1 <0$,
does not occur any more.
}
\end{figure}

\begin{figure}
\includegraphics[width=100mm]{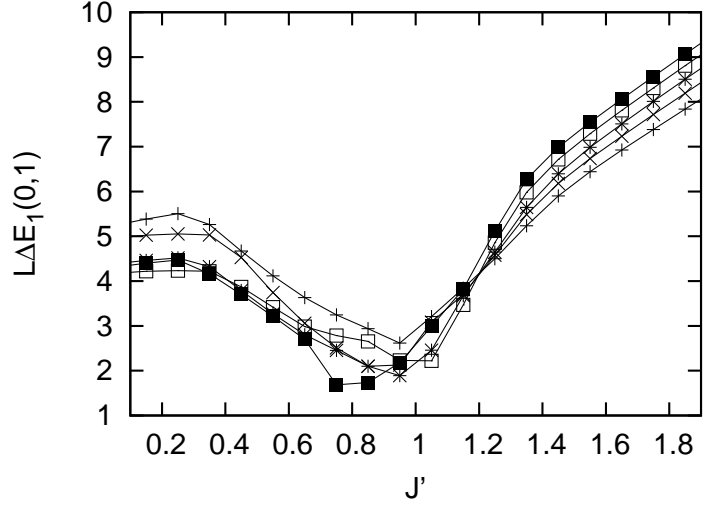}%
\caption{\label{figure5}
The scaled triplet-excitation gap $L \Delta E_1(0,1)$
is plotted for $J_4=0.07$, various $J'$, and 
$N=$
($+$) 16,
($\times$) 20,
($*$) 24,
($\Box$) 28, and
($\blacksquare$) 32.
A continuous transition takes place at $J'_c=1.2$.
}
\end{figure}

\begin{figure}
\includegraphics[width=100mm]{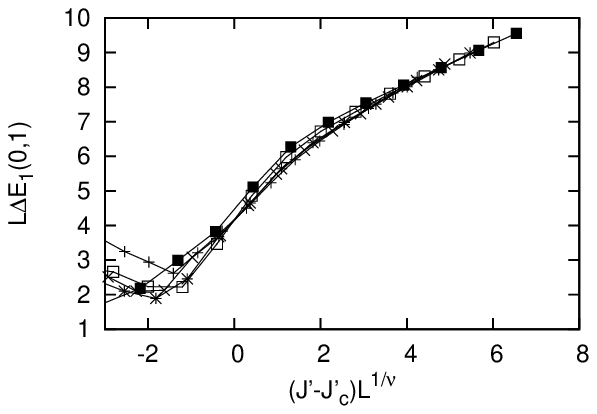}%
\caption{\label{figure6}
The finite-size-scaling plot, $(J'-J'_c)L^{1/\nu}$-$L \Delta E_1(0,1)$,
is shown for
$\nu=0.8$,
$J'_c=1.2$,
$J_4=0.07$, various $J'$, and 
$N=$
($+$) 16,
($\times$) 20,
($*$) 24,
($\Box$) 28, and
($\blacksquare$) 32.
}
\end{figure}

\begin{figure}
\includegraphics[width=100mm]{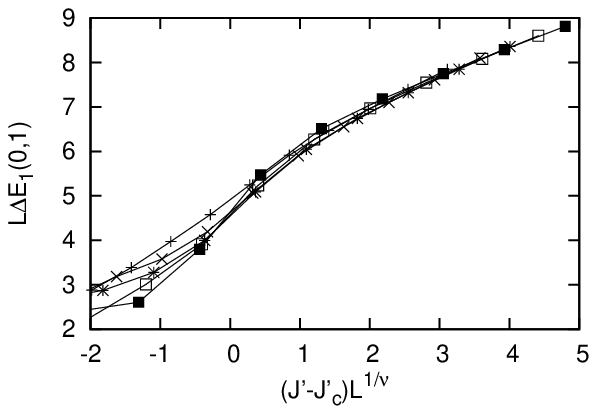}%
\caption{\label{figure7}
The finite-size-scaling plot, $(J'-J'_c)L^{1/\nu}$-$L \Delta E_1(0,1)$,
is shown for
$\nu=0.8$,
$J'_c=1.4$,
$J_4=0.1$, various $J'$, and 
$N=$
($+$) 16,
($\times$) 20,
($*$) 24,
($\Box$) 28, and
($\blacksquare$) 32.
}
\end{figure}

The spatially anisotropic triangular antiferromagnet 
with the ring exchange, Eq. (\ref{Hamiltonian}),
was investigated by means of the numerical diagonalization method.
As the spatial anisotropy $J'$ varies,
the model changes into a variety of systems such as
the one-dimensional, triangular, and square-lattice antiferromagnets
successively.
Taking into account such a geometrical character,
we adopt the screw-boundary condition,
as shown in Fig. \ref{figure2}.


First, we survey the regime without the ring exchange $J_4=0$.
The simulation result indicates that
the intermediate phase appears in $0.65(15)<J'<1.1(1)$.
Our result shows that the VBS phase is robust\cite{Zheng06,Weng06}
against the interchain coupling $J'$.
Second, 
by applying the ring exchange $J_4$,
we suppress the intermediate phase.
Eventually, we attain 
the direct VBS-N\'eel transition, which is under
the current theoretical interest
in the context of the high-temperature superconductivity.
Postulating $z=1$,\cite{Sandvik07,Melko08}
we analyze the simulation data in terms of the finite-size scaling.
Thereby, we estimate the correlation-length critical exponent
as $\nu=0.80(15)$, confirming that the VBS-N\'eel transition is indeed a
critical one.
The exponent is comparable to the preceding Monte Carlo results,
$\nu=0.78(3)$\cite{Sandvik07} and $\nu=0.68(4)$,\cite{Melko08} calculated for the square-lattice antiferromagnet.

Our result provides an evidence that the VBS-N\'eel
transition is critical, realizing 
the deconfinement criticality.
Here, the ring exchange plays a
significant role.
In Ref. \onlinecite{Misguich99}, 
generic types of ring-exchange interactions
are considered in the context of the
Helium adsorbate.
Such an extension may also lead to an improvement
as to the finite-size behavior.
This problem will be addressed in a
future study.




\begin{acknowledgments}
This work was supported by a Grant-in-Aid 
from Monbu-Kagakusho, Japan
(No. 18740234).
\end{acknowledgments}


\end{document}